\pdfoutput=1
\documentclass[twocolumn,showpacs,preprintnumbers,amsmath,amssymb,superscriptaddress]{revtex4-1}
\usepackage{epsfig}
\usepackage{graphicx}
\usepackage{dcolumn}
\usepackage{bm}
\usepackage{float}

\begin{document}

\selectfont

\title{Mitigation of Malicious Attacks on Networks}

\author{Christian M. Schneider}
\email{schnechr@ethz.ch}
\affiliation{Computational Physics, IfB, ETH Zurich, Schafmattstrasse 6, 8093 Zurich, Switzerland}
\author{Andr\'e A. \surname{Moreira}}
\affiliation{Departamento de F\'{\i}sica, Universidade Federal 
do Cear\'a, 60451-970 Fortaleza, Cear\'a, Brazil}
\author{Jos\'e S. Andrade Jr.}
\affiliation{Computational Physics, IfB, ETH Zurich, Schafmattstrasse 6, 8093 Zurich, Switzerland}
\affiliation{Departamento de F\'{\i}sica, Universidade Federal
do Cear\'a, 60451-970 Fortaleza, Cear\'a, Brazil}
\author{Shlomo \surname{Havlin}}
\affiliation{Minerva Center and Department of Physics, Bar-Ilan University, 52900 Ramat-Gan, Israel}
\author{Hans J. Herrmann}
\affiliation{Computational Physics, IfB, ETH Zurich, Schafmattstrasse 6, 8093 Zurich, Switzerland}
\affiliation{Departamento de F\'{\i}sica, Universidade Federal
do Cear\'a, 60451-970 Fortaleza, Cear\'a, Brazil}

\date{\today}
 
\begin{abstract}
Terrorist attacks on transportation networks have traumatized modern societies. With a single blast, it has become possible to paralyze airline traffic, electric power supply, ground transportation or Internet communication. How and at which cost can one restructure the network such that it will become more robust against a malicious attack? We introduce a unique measure for robustness and use it to devise a method to mitigate economically and efficiently this risk. We demonstrate its efficiency on the European electricity system and on the Internet as well as on complex networks models. We show that with small changes in the network structure (low cost) the robustness of diverse networks can be improved dramatically while their functionality remains unchanged. Our results are useful not only for improving significantly with low cost the robustness of existing infrastructures but also for designing economically robust network systems.
\end{abstract}

%%%%PACS e Keywords
 \pacs{64.60.ah, %Percolation
       64.60.aq, %Networks
       89.75.Da, %Systems obeying scaling laws
       89.75.Fb, %Structures and organization in complex systems
       89.75.Hc %Networks and genealogical trees
       \\~\\
       PNAS March 8, 2011 vol. 108 no. 10 3838-3841\\
       http://www.pnas.org/content/108/10/3838.full\\
       doi:10.1073/pnas.1009440108
} 
 
\keywords{percolation | power grid}

\maketitle

The vulnerability of modern infrastructures stems from their network structure having very high degree of interconnectedness which makes the system resilient against random attacks but extremely vulnerable to targeted raids {\cite{Barabasi,Watts,barab,computernetworks,computernetworks1,socialnetworks,barabreview,socialnetworks1,Caldarelli02,computernetworks2,Dorogovtsev,Newman,powernetworks,Valente,Caldarelli,auto,Hooyberghs10}}. We developed an efficient mitigation method and discovered that with relatively minor modifications in the topology of a given network and without increasing the overall length of connections, it is possible to mitigate considerably the danger of malicious attacks. Our efficient mitigation method against malicious attacks is based on developing and introducing a unique measure for robustness. We show that the common measure for robustness of networks in terms of the critical fraction of attacks at which the system completely collapses, the percolation threshold, may not be useful in many realistic cases. This measure, for example, ignores situations in which the network suffers a significant damage, but still keeps its integrity. {Besides the percolation threshold, there are other robustness measures based, for example, on the shortest path \cite{Frank70,Latora01,Sydney10} or on the graph spectrum \cite{Fiedler73}. They are, however, less frequently used for being too complex or less intuitive. In contrast, our unique robustness measure, which considers the size of the largest component during all possible malicious attacks, is as simple as possible and only as complex as necessary. Due to the ample range of our definition of robustness, we can assure that our process of reconstructing networks maintains the infrastructure as operative as possible, even before collapsing.}

\begin{figure*}
 \begin{minipage}[c]{.45\textwidth}
  \includegraphics[height=.96\textwidth,angle = 0]{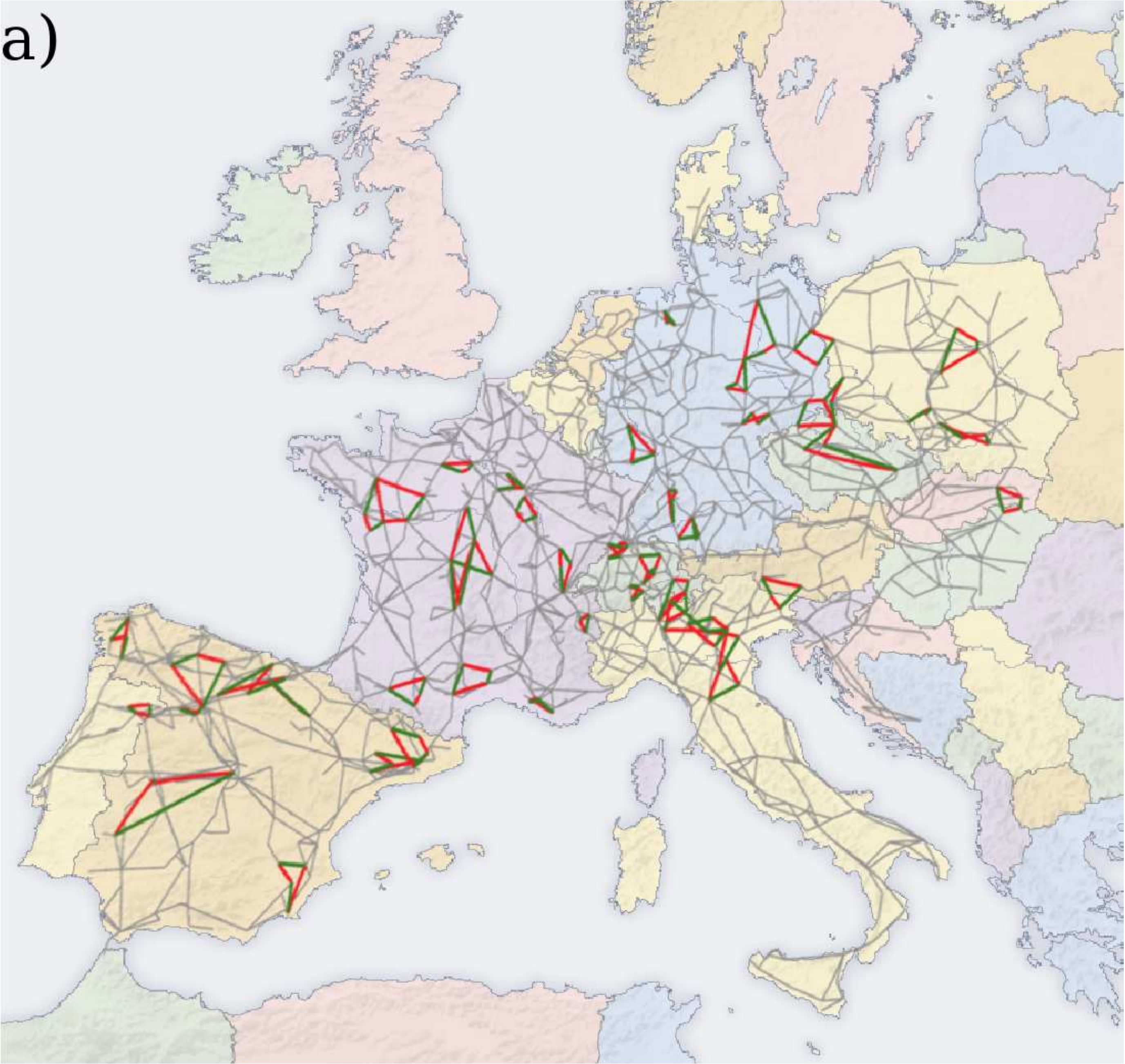}
 \end{minipage}
 \begin{minipage}[c]{.45\textwidth}
  \includegraphics[height=.96\textwidth,angle = 0]{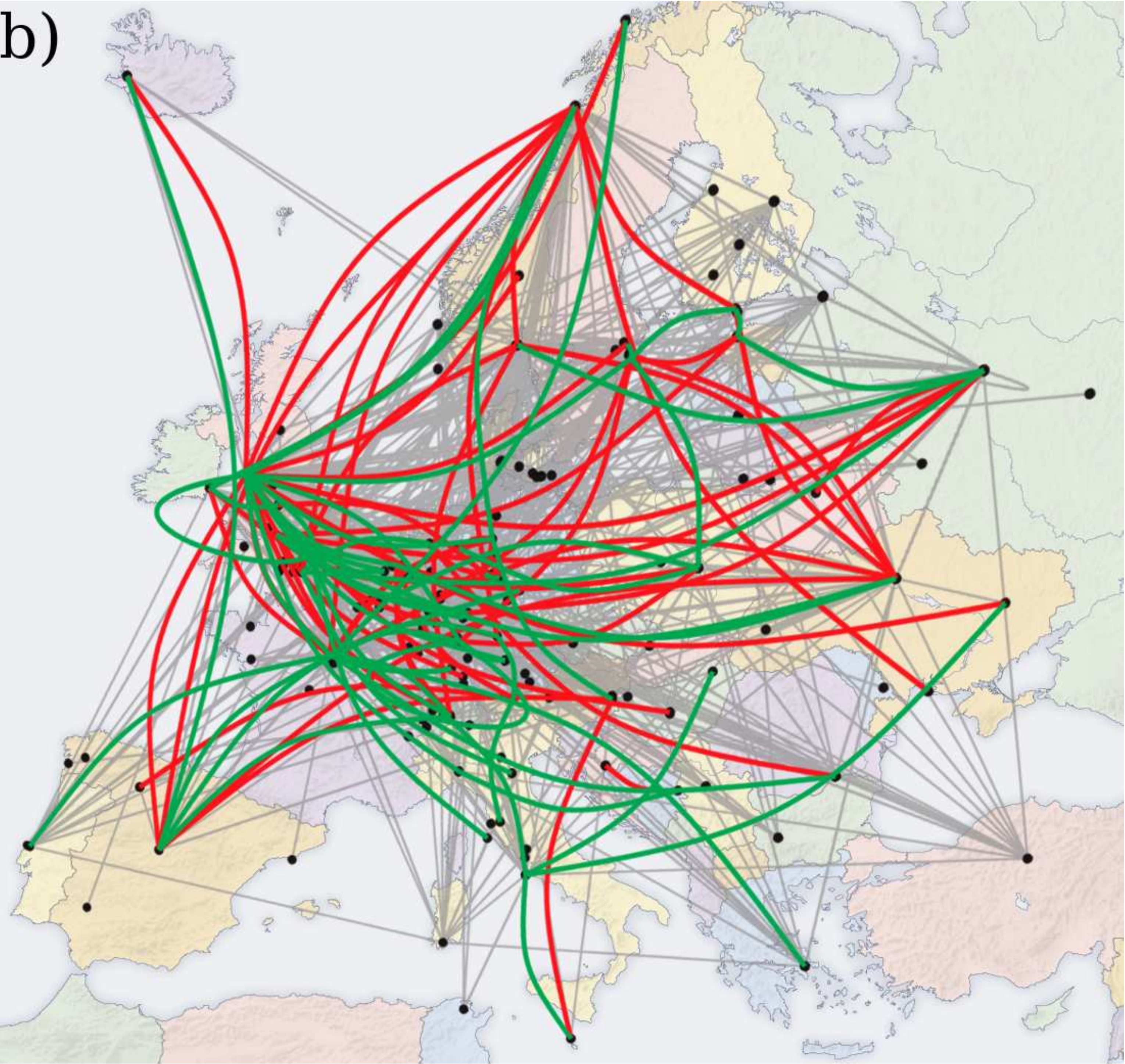}
 \end{minipage}\\
 \begin{minipage}[c]{.45\textwidth}
  \includegraphics[height=.96\textwidth,angle = 0]{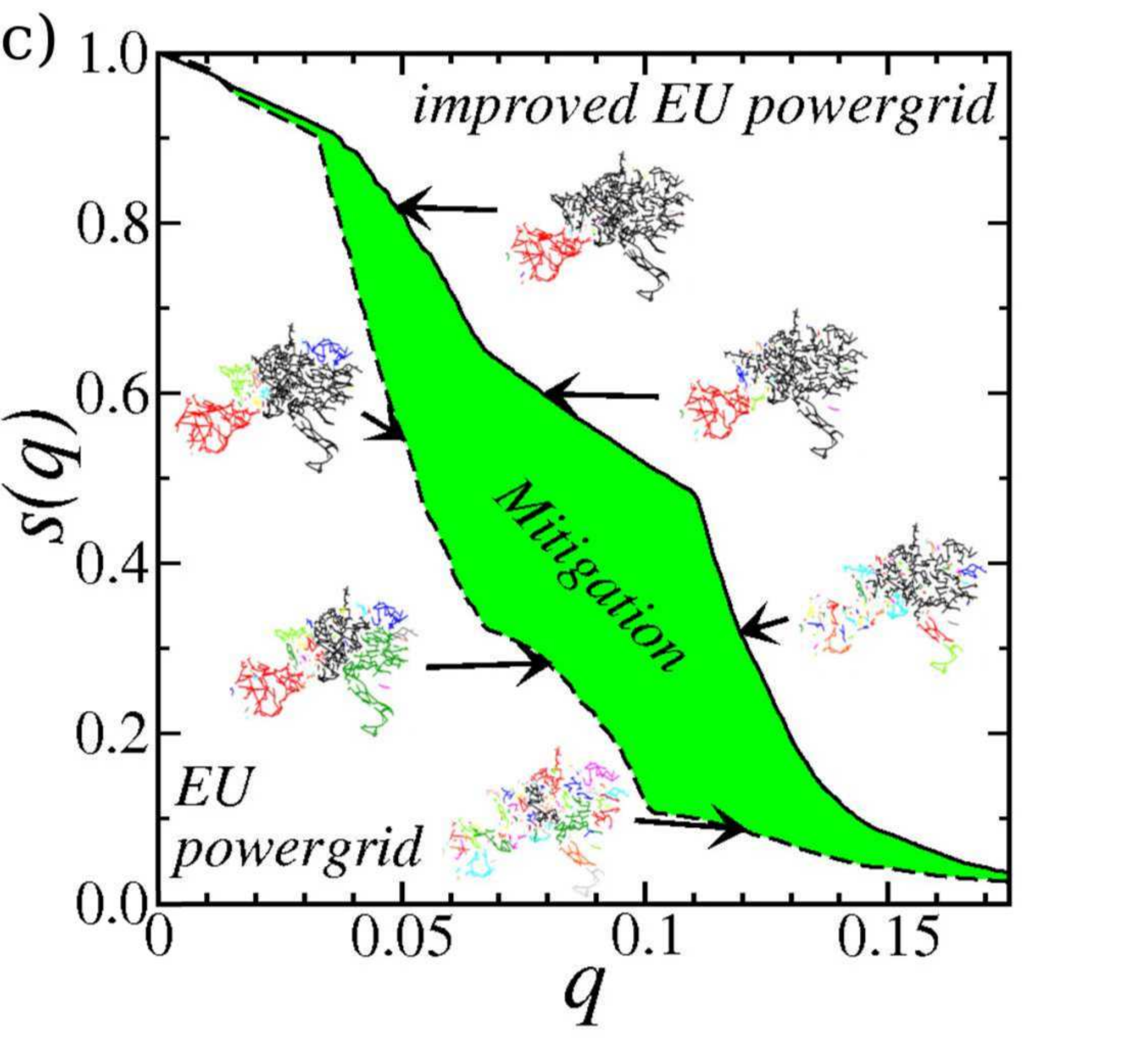}
 \end{minipage}
 \begin{minipage}[c]{.45\textwidth}
  \includegraphics[height=.96\textwidth,angle = 0]{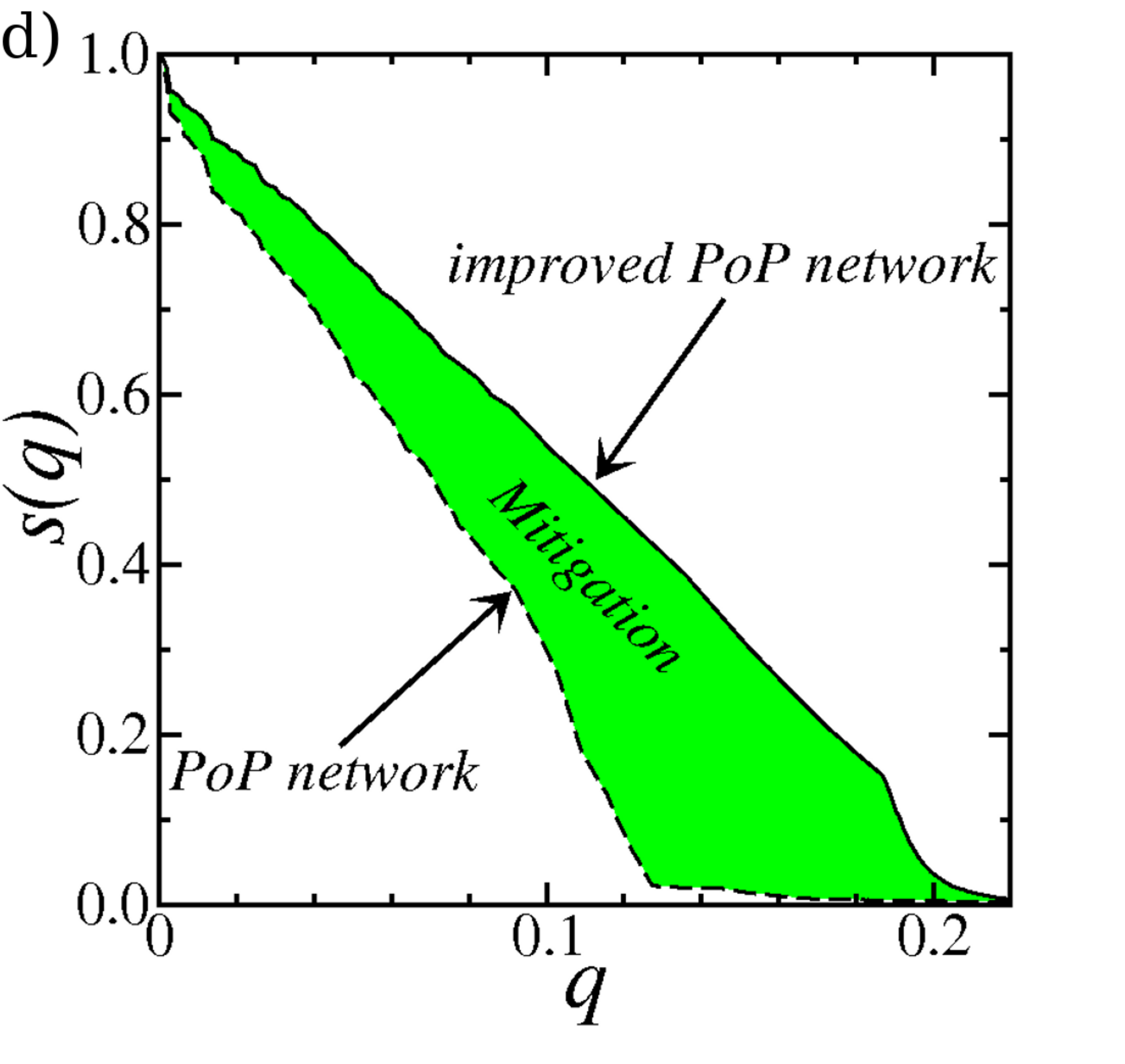}
 \end{minipage}
\caption{Mitigation of malicious attacks on the power supply system in Europe and the global Internet at the level of service providers. In (a) we show the EU power grid with $N = 1254$ generators and $M = 1811$ power lines \cite{power1} and in (b) the Internet with $N = 1098$ service providers and $M = 6089$ connection among them, where only the European part is shown \cite{Pajek}. The red edges correspond to the $5\%$ connections that we suggest to replace by the green ones. {A detailed description of the chosen edges is given in the SI.} The network fragmentation under a malicious attack is shown for (c) EU power generators and for (d) PoP. The dashed lines in (c) and (d) corresponds to the size of the largest component in each original system and the solid lines to {typical} redesigned networks after changing $5\%$ of the connections. The green areas give the mitigation of malicious attack, which correspond to improving robustness by $45\%$ for the EU power grid and $55\%$ for the PoP.}
\label{fig:topology}
\end{figure*}

\begin{figure*}
 \begin{minipage}[c]{.45\textwidth}
  \includegraphics[width=6.4cm,angle = -90]{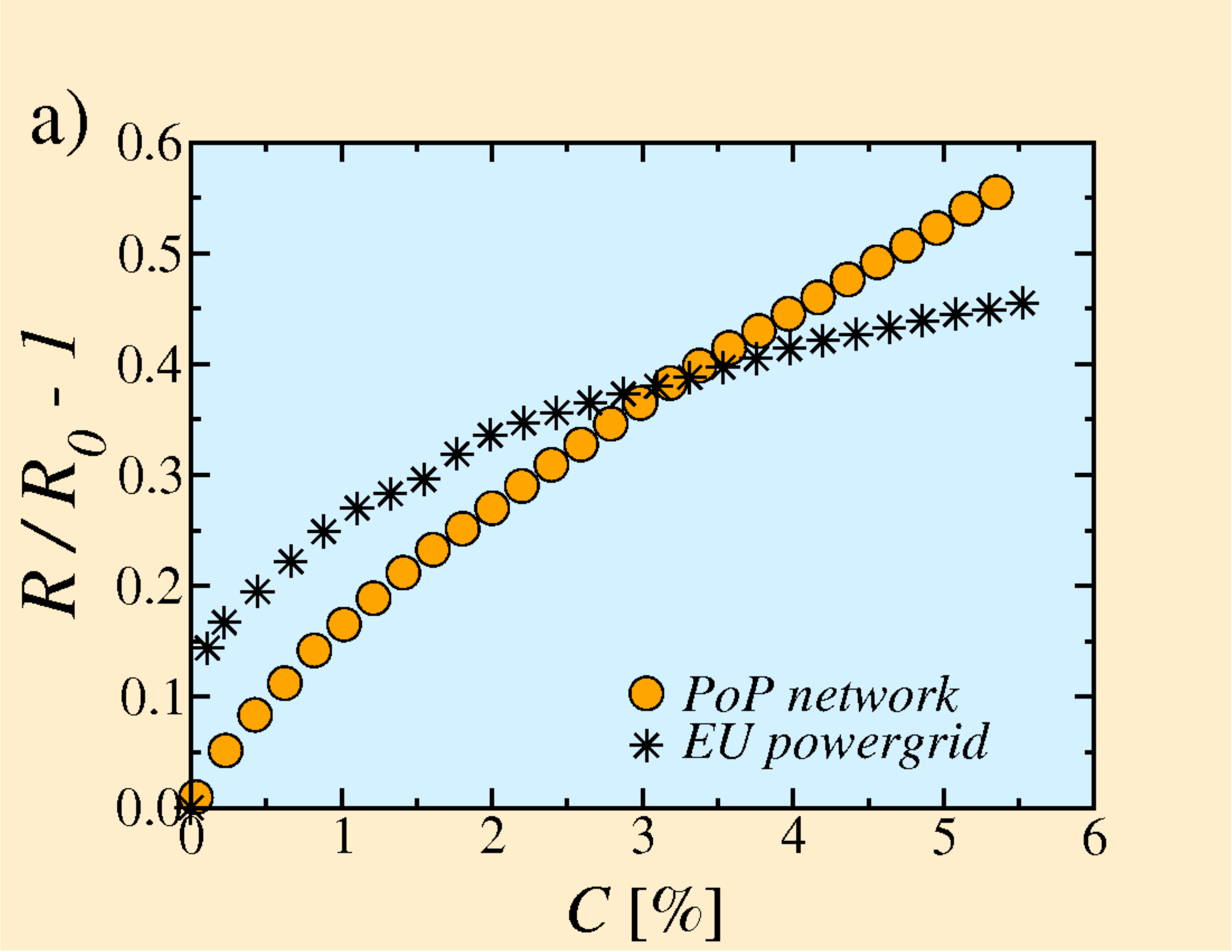}
 \end{minipage}
 \begin{minipage}[c]{.45\textwidth}
  \includegraphics[width=6.4cm,angle = -90]{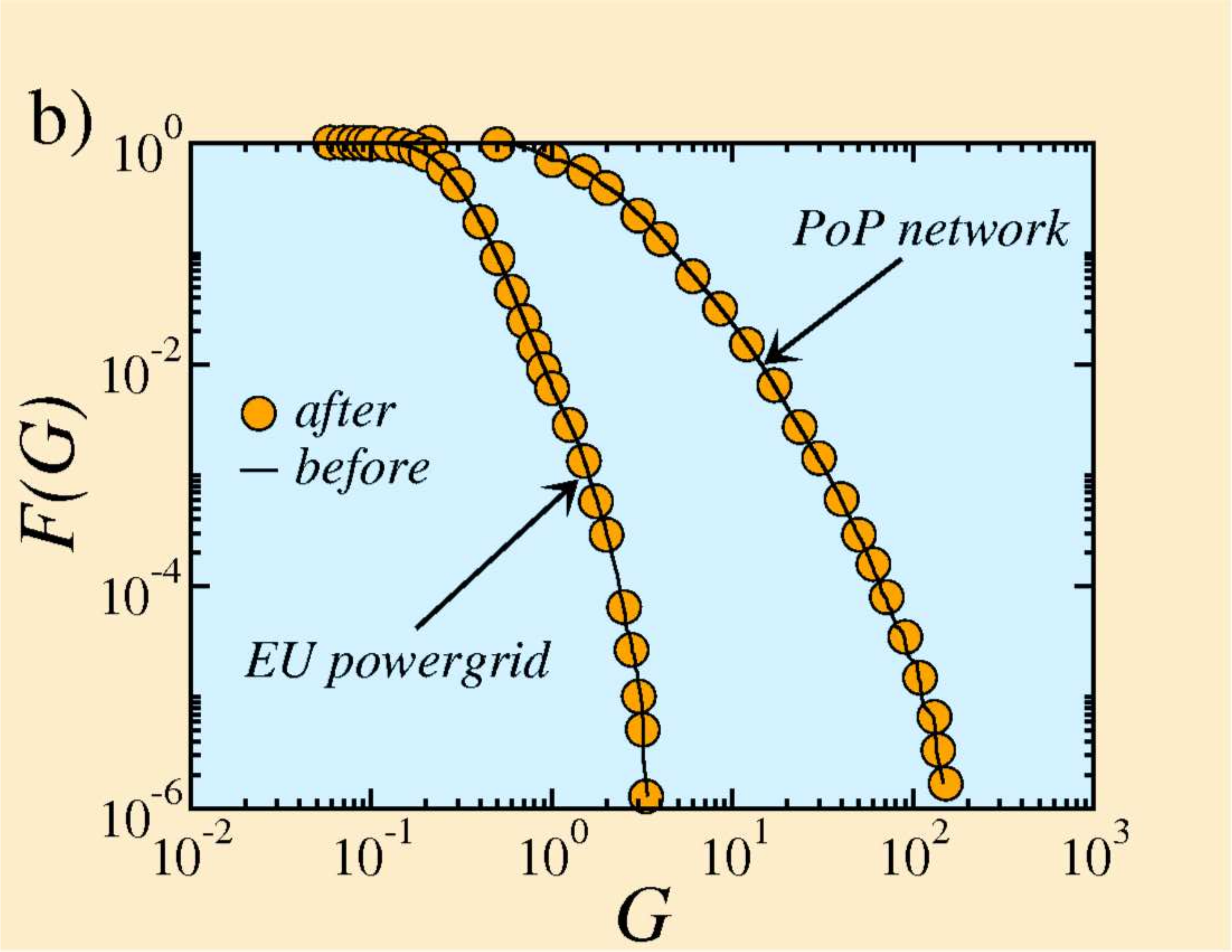}
 \end{minipage}
\caption{Demonstration that small changes have a large impact on the robustness while the functionality of the networks remains. a) Improvement of robustness $R$ as a function of the fraction of changed links for both networks, where $R_0$ is the original robustness. In the case of the EU power grid, we find that changing only two connections increases the robustness by $15\%$. When changing $2\%$ of the links, the robustness of the EU power grid improves by $35\%$ and the Internet by $25\%$. b) The cumulative conductance distribution $F(G)$ versus the conductance $G$ for both networks before and after the changes. Conductances between two nodes are measured for all pairs of nodes, assuming that each link in the network has unitary conductance. Both curves are nearly identical, which means that the transport properties, i.e., the functionalities of the improved networks are very close to the original ones.}
\label{fig:topology1}
\end{figure*}

\section{Model}
\subsection{Modeling attack on infrastructures}
We begin by demonstrating the efficiency of our unique approach to improve the performance of two of the most fragile, but critical infrastructures, namely, the power supply system in Europe \cite{power1} as well as the global Internet at the level of service providers, the so-called Point of Presence (PoP) \cite{PoP}. The breakdown of any of these networks would constitute a major disaster due to the strong dependency of modern society on electrical power and Internet. In Figs.~\ref{fig:topology}a and \ref{fig:topology}b we show the backbone of the European power grid and the location of the European PoP and their respective vulnerability in Figs.~\ref{fig:topology}c and \ref{fig:topology}d. The dotted lines in Figs.~\ref{fig:topology}c and \ref{fig:topology}d represent the size of the largest connected component of the networks
after a fraction $q$ of the most connected nodes have been removed. {Instead of using the static approach to find the $q$ most connected nodes at the beginning of the attack, we use a dynamical approach. In this case the degrees are recalculated during the attack, which corresponds to a more harmful strategy \cite{holme}.} As a consequence, in their current structure, the shutdown of only $10\%$ of the power stations and a cut of $12\%$ of PoP would affect $90\%$ of the network integrity. In order to avoid such a dramatic breakdown and reduce the fragility of these networks, here we propose a strategy to exchange only a small number of power lines or cables without increasing the total length of the links and the number of links of each node. These small local changes not only mitigate the efficiency of malicious attacks, but at the same time preserve the
functionality of the system. In Figs.~\ref{fig:topology}c and \ref{fig:topology}d the robustness of the original networks are given by the areas under the dashed curves, while the areas under the solid lines correspond to the robustness of the improved networks. Therefore, the green areas in Figs.~\ref{fig:topology}c and \ref{fig:topology}d demonstrate the significant improvement of the resilience of the network for any fraction $q$ of attack. This means that terrorists would cause less damage or they would have to attack more power stations, and hackers would have to attack more PoP in order to significantly damage the system.

\subsection{Introducing the unique robustness measure}
Next, we describe in detail our methodology. Usually robustness is measured by the value of $q_c$, the critical fraction of attacks at which the network completely collapses \cite{holme}. This measure ignores situations in which the network suffers a big damage without completely collapsing. We thus propose here a unique measure which considers the size of the largest component during {\it all possible} malicious attacks. Malicious raids often consist of a certain fraction $q$ of hits and we want to assure that our process of reconstructing networks will keep the infrastructure as operative as possible, even
before collapsing. Our unique robustness measure $R$, is thus defined as,
\begin{eqnarray}
 R = \frac{1}{N} \sum_{Q = 1}^{N} s(Q)~,
\end{eqnarray}
where $N$ is the number of nodes in the network and $s(Q)$ is the fraction of nodes in the largest connected cluster after removing $Q = q N$ nodes. The normalization factor $1/N$ ensures that the robustness of networks with different sizes can be compared. The range of possible $R$ values is between $1/N$ and $0.5$, where these limits correspond, respectively, to a star network and a fully connected graph.

\subsection{Constraints for improving networks}
For a given network, the robustness could be enhanced in many ways. Adding links without any restrictions until the network is fully connected would be an obvious one. However, for practical purposes, this option can be useless since, for example, the installation of power lines between each pair of power plants would skyrocket costs and transmission losses. 
{By associating costs to each link of the network, we must seek for a reconstruction solution that minimizes the total cost of the changes. We also assume that changing the degree of a node can be particularly more expensive than changing edges. These two assumptions suggest keeping invariant the number of links and the degree of each node.}
Under these constraints, we propose the following algorithm to mitigate malicious attacks. In the original network we swap the connections of two randomly chosen edges, that is, the edges $e_{ij}$ and $e_{kl}$, which connect node $i$ with node $j$, and node $k$ with node $l$, respectively, become $e_{ik}$ and $e_{jl}$ \cite{Maslov02}, only if the robustness of the network is increased, i.e., $R_\mathrm{new} > R_\mathrm{old}$. {Note that a change of the network usually leads to an adjustment in the attack sequence.} We then repeat this procedure with another randomly chosen pair of edges until no further substantial improvement is achieved for a given large number of consecutive swapping trials. In Fig.~1 of the SI we show numerical tests indicating that the algorithm can indeed yield close to optimal robustness.
As described so far, our algorithm can be used to improve a network against malicious attacks while conserving the number of links per node. Nevertheless, for real networks with economical constraints, this conservation of degree is not enough since the cost, like the total length of links, can not be exceedingly large and also the number of changes should remain small. Therefore, for reconstructing the EU power grid and the worldwide PoP, we use an additional condition that the swap of two links is only accepted if the total length (geographically calculated) of edges does not increase and the robustness is increased by more than a certain value.

\begin{figure*}
 \begin{minipage}[c]{.45\textwidth}
  \includegraphics[width=6.4cm,angle = -90]{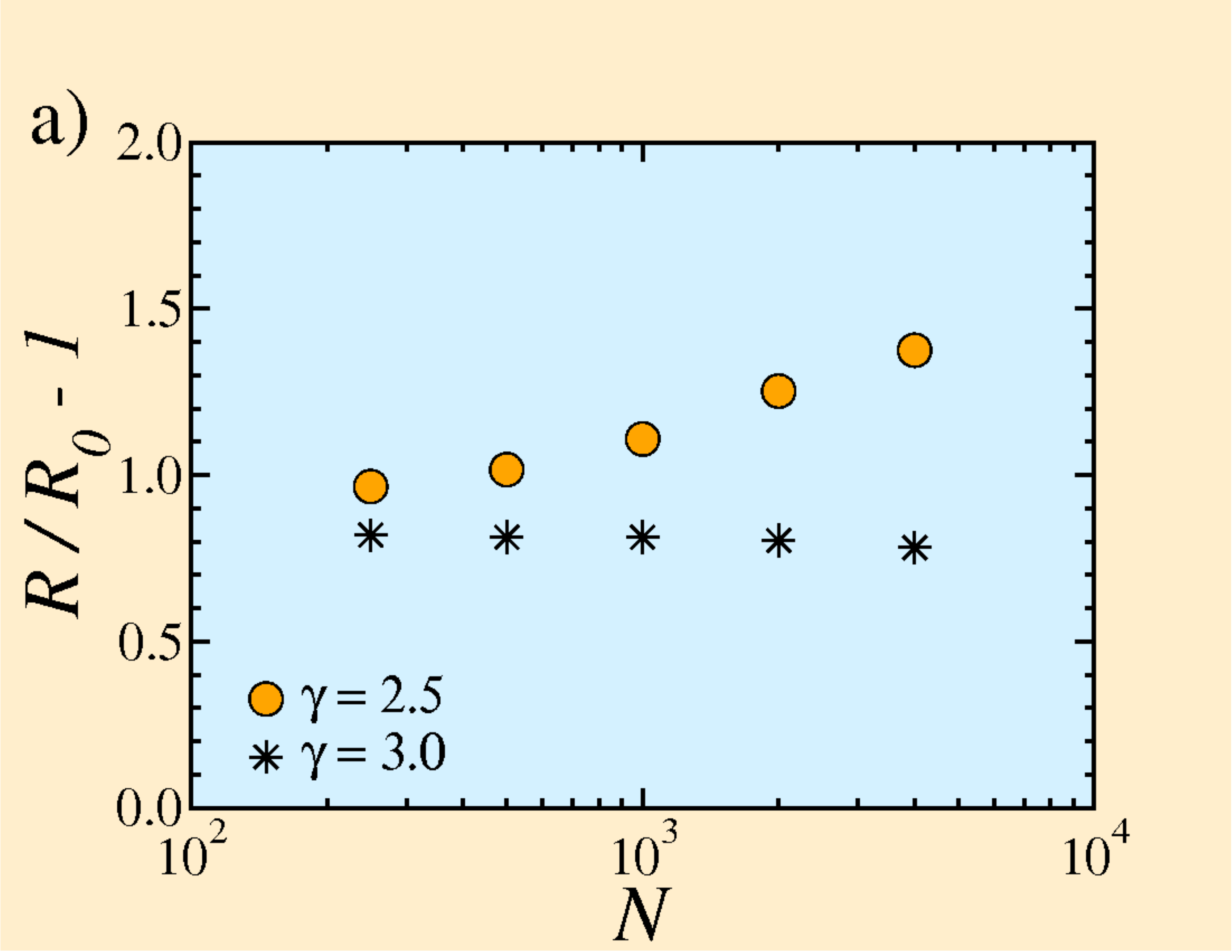}
 \end{minipage}
 \begin{minipage}[c]{.45\textwidth}
  \includegraphics[width=6.4cm,angle = -90]{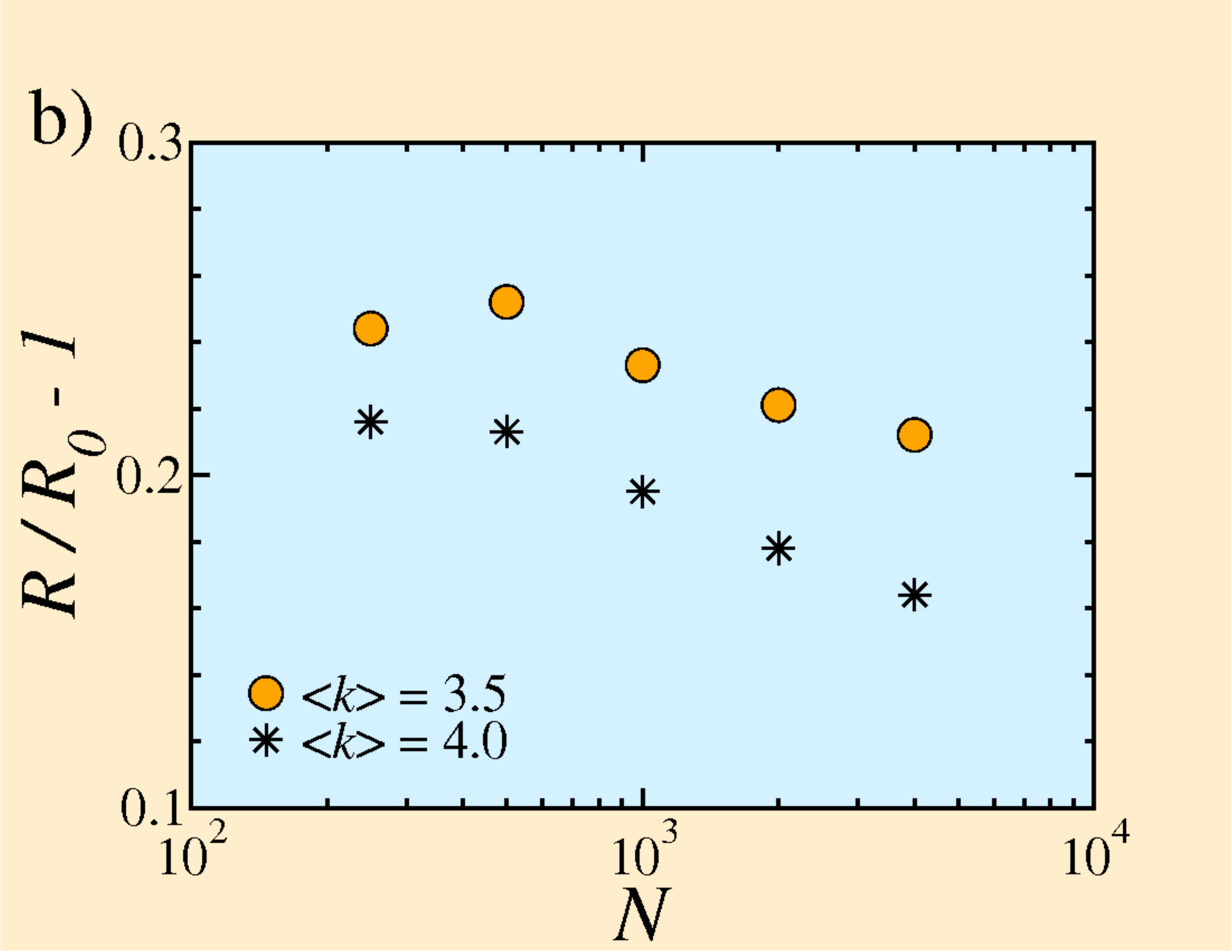} 
 \end{minipage}
\caption{Validation that one can design robust networks regardless of the degree distribution and the system size. The relative robustness improvement $R/R_0 - 1$ vs network size $N$ for (a) scale-free networks with degree exponent $\gamma = 2.5$ and $3$ and (b) Erd\H{o}s-R\'enyi networks with $\langle k \rangle = 3.5$ and $4$. Starting from a given network, we swap two randomly chosen connections, that is, $e_{ij}$, which connects node $i$ with node $j$, and $e_{kl}$ become $e_{ik}$ and $e_{jl}$, only if the robustness of the network is increased. This procedure is repeated until during the last $10000$ attempts no further improvement could be achieved. Note that the swapping keeps the degree of each node unchanged. {Results are averaged over at least five independent initial networks. We do not show error bars, since they are smaller than the symbol sizes.}}
\label{fig:size}
\end{figure*}

\section{Results}
\subsection{Improving existing infrastructures}

Figure~\ref{fig:topology1}a shows that, despite these strong constraints, the robustness $R$ can be increased by $55\%$ for PoP and $45\%$ for the EU grid with only $5.5\%$ of link changes and by $34\%$ and $27\%$, respectively, with only $2\%$. Interestingly, although the robustness is clearly improved, we observe that the percolation threshold $q_c$ remains practically the same for both networks, justifying our unique definition for the measure $R$ as a robustness criterion. More strikingly, the conductance distribution \cite{conductance}, which is a useful measure for the functionality of the network, also does not change (see Fig.~\ref{fig:topology1}b). This suggests that our optimized network is not only more robust against malicious attacks, but also does not increase the total length of connections without any loss of functionality.\\
\begin{figure*}
 \begin{minipage}[c]{.45\textwidth}
  \includegraphics[width=6.4cm,angle = -90]{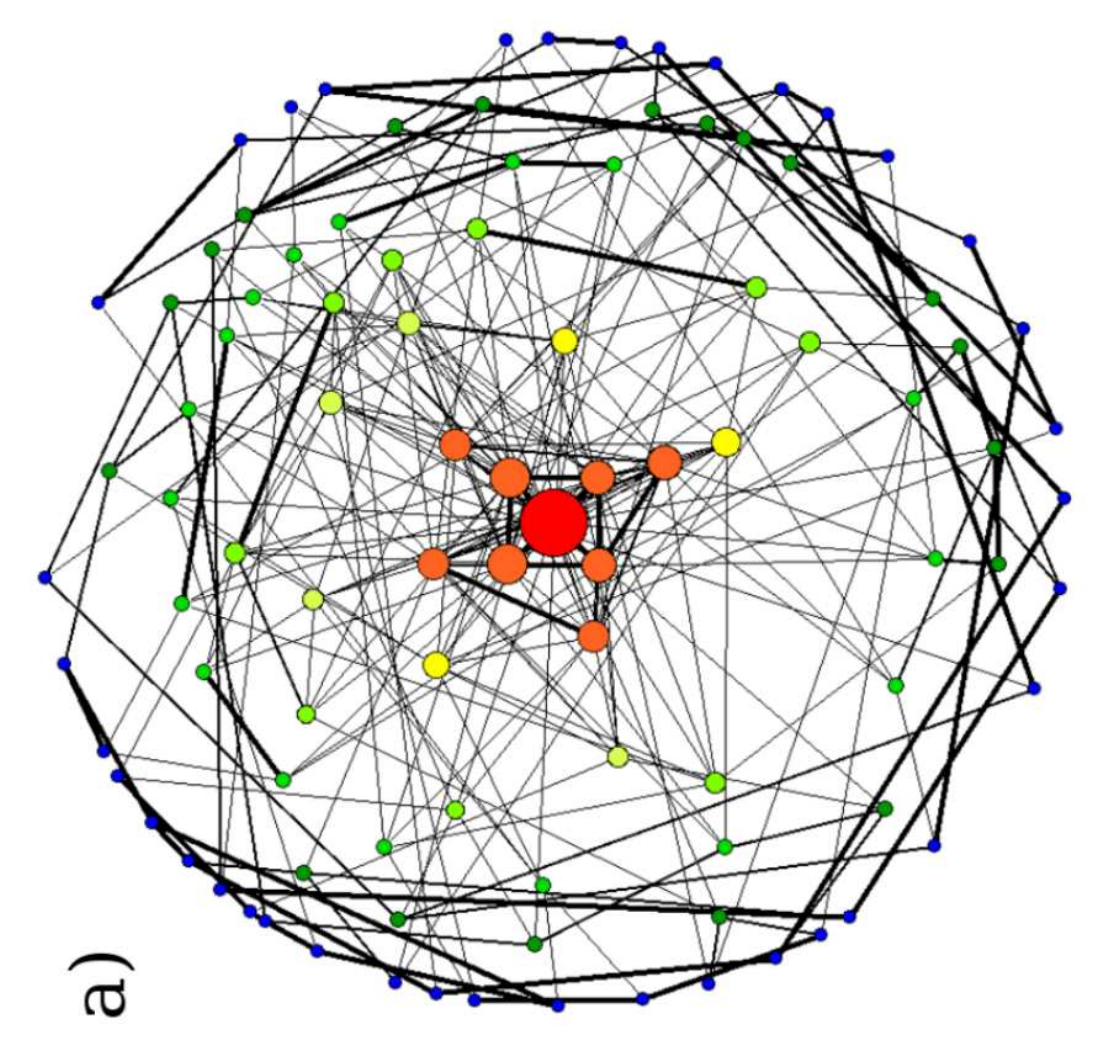}
 \end{minipage}
 \begin{minipage}[c]{.45\textwidth}
  \includegraphics[width=6.4cm,angle = -90]{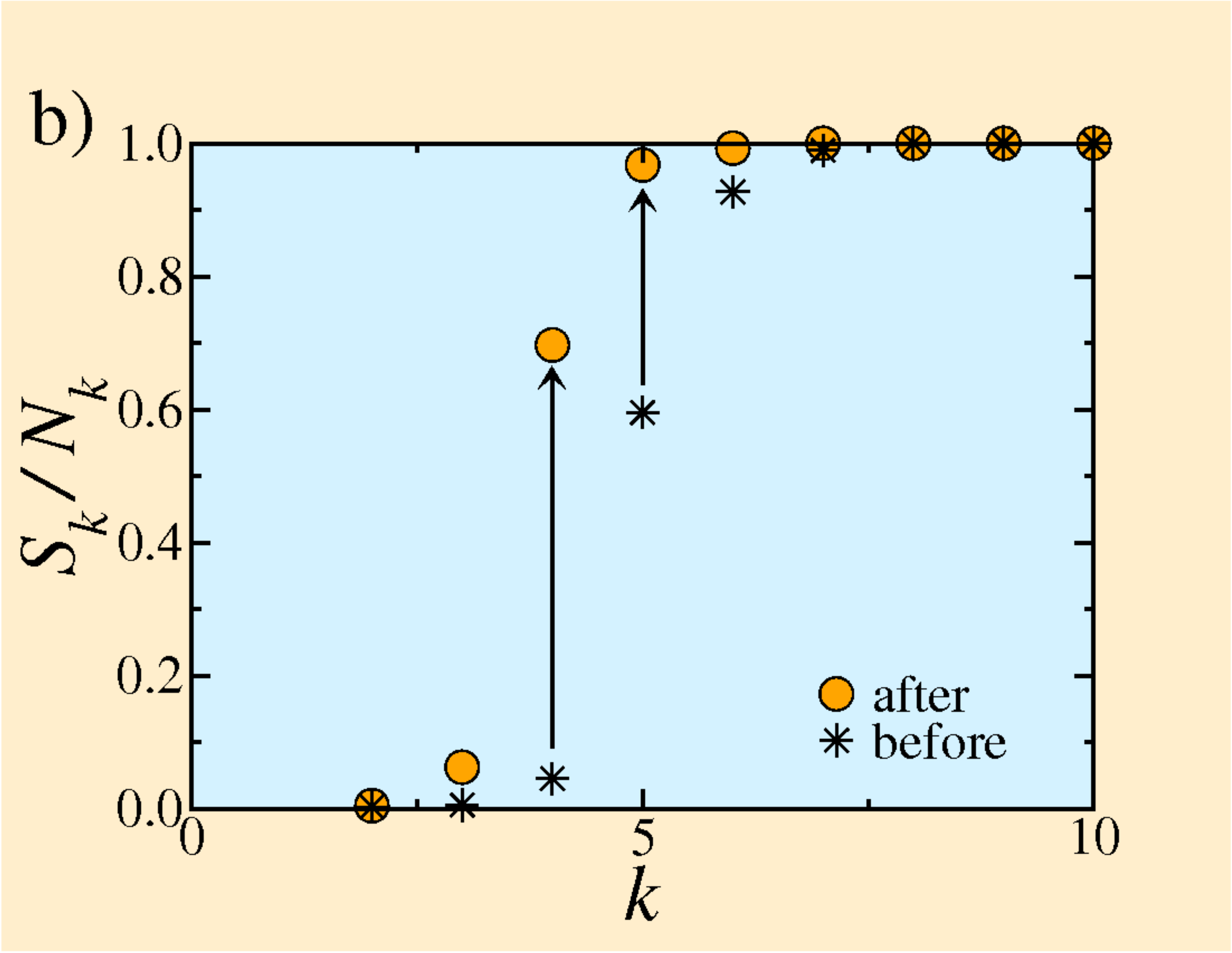}
 \end{minipage}
\caption{Visualization of the novel onion-like topology of robust networks.
a) The onion-like topology of a robust scale-free network with $N = 100$ nodes, $M = 300$ edges and a degree distribution $P(k) \sim k^{-2.5}$. The sizes of the nodes are proportional to their degree, and nodes with similar degree have the same color. Edges between nodes with equal degree and the fully connected core are highlighted. In onion-like networks nearly each pair of nodes with equal degree $k$ is connected by a path that does not contain nodes of higher degree. b) Fraction of nodes with degree $k$ that are connected through nodes with a degree smaller or equal to $k$ for scale-free networks with $\gamma = 2.5$ and $N = 4000$.}  \label{fig:onion}
\end{figure*}

\subsection{Designing robust networks}
The success of this method in reconstructing real networks to improve robustness at low cost and small effort leads us to the following question: Can we apply our algorithm to design new highly robust networks against malicious attacks? In this case, since we build the network from the beginning, the number of changes should not represent any limitation, since we are dealing with only a computational problem. For designing, the only constraint which remains is the invariance of the degree distribution. Here we study both artificial scale-free \cite{static} and Erd\H{o}s-R\'enyi networks \cite{Erdos}. In Fig.~\ref{fig:size} we show how the robustness depends on the system size for designed scale-free networks with degree
distribution $P(k) \sim k^{-\gamma}$, with $\gamma = 2.5$ and $3$, and Erd\H{o}s-R\'enyi networks with average degree $\langle k \rangle = 3.5$ and $4$. One can see that our method is also very efficient in designing robust networks.

While the most robust network structure for a given degree distribution is virtually impossible to determine, our study reveals that all networks investigated can be improved significantly (see Fig.~\ref{fig:size} and Fig.~2 in SI). Moreover, as shown in Fig.~\ref{fig:onion}a, the robust networks we obtain clearly share a common and unique ``onion-like'' structure consisting of a core of highly connected nodes hierarchically surrounded by rings of nodes with decreasing degree. To quantitatively test our observation, we calculate the maximal number of nodes $S_k$ with degree $k$ which are connected through nodes with a degree smaller or equal to $k$. As shown in Fig.~\ref{fig:onion}b, paths between nodes of equal degree, which are not passing through nodes with higher degree, emerge in the robust networks. Although at a first glance onion-like networks might look similar to high assortative networks, the later ones are different and can be significantly more fragile (see Fig.~3 in SI). We also find that onion-like networks are also robust against other kinds of targeted attacks such on high betweenness nodes \cite{holme} (see Fig.~4 in SI). {The last topological properties we study are the average shortest path length between two nodes, $l$, and the diameter, $d$, corresponding to the maximal distance between any pair of nodes \cite{barabreview}. Counter intuitively, $l$ and $d$ do not decrease after the optimization, but slightly increase. Nevertheless, it seems that both values grow not faster than logarithmically with the system size $N$. (see Fig.~5 in SI)\\}

\section{Discussion}
In summary, we have introduced a unique measure for robustness of networks and used this measure to develop a method that significantly improves, with low cost, their robustness against malicious attacks. Our approach has been found to be successfully useful as demonstrated on two real network systems, the European power grid of stations and the Internet. Our results show that with a reasonably economical effort, significant gains can be achieved for their robustness while conserving the nodes degrees and the total length of power lines or cables. In the case of designing scale-free networks, a unique ``onion-like'' topology characterizing robust networks is revealed. This insight enables to design robust networks with a prescribed degree distribution. The applications of our results are imminent on one hand to guide the improvement of existing networks but also serve on the other hand to design future infrastructures with improved robustness.

\begin{acknowledgments}
We thank T. Mihaljev for useful discussions, and Y. Shavitt and N. Zilberman for providing the Point of Presence Internet data. We acknowledge financial support from the ETH Competence Center ``Coping with Crises in Complex Socio-Economic Systems'' (CCSS) through ETH Research Grant CH1-01-08-2. S.H. acknowledges support from the Israel
Science Foundation, ONR, DTRA and the Epiwork EU project. A.A.M. and J.S.A would like to thank CNPq, CAPES, FUNCAP, and FINEP for financial support.
\end{acknowledgments}

\renewcommand{\figurename}{Supporting Figure}
\renewcommand{\textfraction}{0.0}
\renewcommand{\topfraction}{1.0}
\renewcommand{\bottomfraction}{1.0}
\addtocounter{figure}{-4}
\newpage

{
\bf Supporting Information Appendix
}
~\\
~\\
A. Detailed description of our algorithm for Fig. 1: Our algorithm starts with calculating the robustness of the original PoP (power supply system). Therefore, the robustness $R$ of $2000$($1000$) independent attacks based on the adaptive calculation of the highest degree nodes are calculated. The average robustness value is assigned to the initial robustness $R_\text{old}$.\\
To improve the network's robustness, two nodes $i$ and $j$, and two of their neighbors $k$ and $l$, are chosen randomly. If the swap of the neighbors, $l$ becomes neighbor of $i$ and $k$ becomes neighbor of $j$, neither creates self-connections nor double-connections, the geographical lengths of the edges $e_{il}$ and $e_{jk}$ is determined. Only if the sum of these lengths is shorter than the sum of the lengths of $e_{ik}$ and $e_{jl}$, the robustness of the new network $R_\text{new}$ is calculated, again averaged over $2000$($1000$) independent attacks. The swap of the neighbors is accepted, only if it would increase the robustness significantly $R_\text{new} > R_\text{old} + \delta_R$, whereas the threshold $\delta_R$ is arbitrary set to $\delta_R = 0.0006$($0.001$). Note that the threshold should be so large that nearly every change is rejected. After testing $10^6$($10^8$) independent swaps, the threshold is reduced by a arbitrary factor of $0.8$.\\
Then $10^6$($10^8$) more swaps are performed with the new threshold, before the threshold is decreased again by a factor of $0.8$. This loop is repeated $10$ times and finally, after $10^7$($10^9$) tested swaps, the improved networks, shown in Fig. 1, are obtained.\\
Note that the including of the decreasing threshold ensures that changes with the largest impact, are performed first.\\

B. Optimal test: In order to test if our algorithm identifies an optimal or near-to-optimal solution, we applied our procedure to three different types of artificial networks, having all the same number of links, but different degree distributions. The first type is the scale-free network $\lbrack$27$\rbrack$, the second is the Erd\H{o}s-R\'enyi $\lbrack$28$\rbrack$ network and the third a random regular network having a fixed degree. While the most robust network against malicious attacks for a given degree distribution is unknown, the most robust network for a given number of edges is a network in which all nodes have the same degree. Therefore, to test our model we will only impose in our algorithm the constraint of conserving the total number of links, but allowing changes of the degree distribution. In this way, the original swapping mechanism of two connections in the algorithm is replaced by the exchange of a given edge by another one which connects two randomly, but not connected nodes. We find that the robustness of the final networks are practically indistinguishable, as shown in Supporting Figure 1a. Not only their robustness is similar, but also the obtained degree distributions of the networks converge to a delta function around $\langle k \rangle$ (see Supporting Figure 1b). These results represent strong indication that our algorithm efficiently finds the structure very close to the most robust network. Simulations with different initial realizations of distinct networks types also converged to similar final states in both robustness and degree distribution.\\

C. Model networks: In Supporting Figs.~2a and 2c we show the robustness of scale-free and Erd\H{o}s-R\'enyi network models. The robustness of the model networks are given by the areas under the dashed curves, while the areas under the solid lines correspond to the robustness of the improved networks. Therefore, the green areas indicate the improvement of the resilience of the network for any fraction $q$ of attack. As shown in Supporting Figs.~2b and 2d, the overall behavior of the conductance distribution $F(G)$, which is a useful measure for the transport functionality of the network, also does not change.\\

D. Assortative networks: In Supporting Fig.~3 we show the result of our method when applied to a high assortative network ($r=0.65$) that does not display onion-like structure. Although these two networks have the same degree distribution, their topology is quite different. The percolation thresholds $q_c$ for both networks are close, but the onion-like network is definitely more robust. It is obvious that onion-like and assortativity are distinct properties and that our new robustness measure is significantly more adequate compared to the classical measure $q_c$.\\

E. High betweenness attack: Instead of removing the most connected nodes from a network, other attack strategies can also be used. For example, one of the most harmful strategies is the so-called high-betweenness based adaptive attack$\lbrack$14$\rbrack$. In this case the nodes are removed according to their betweenness in the network after removing each node. In Supporting Figure 4a and 4b the robustness against this type of attack is shown for scale-free and Erd\H{o}s-R\'enyi networks. Note that although the networks are optimized against high degree-based attack, the designed networks become also significantly more resilient to high-betweenness adaptive attack.\\

{F. Properties of onion-like networks: In Supporting Figs.~5a and 5b we show the average shortest path length and the diameter of onion-like networks obtained from scale-free and Erd\H{o}s-R\'enyi network models. While both properties remain similar for Erd\H{o}s-R\'enyi and the improved network, both properties increase for onion-like networks starting from scale-free networks. Nevertheless, the diameter and the average shortest path length increase not faster than logarithmic with system size for onion-like networks.}

\begin{figure*}[h]
\includegraphics[width=6.cm,angle = -90]{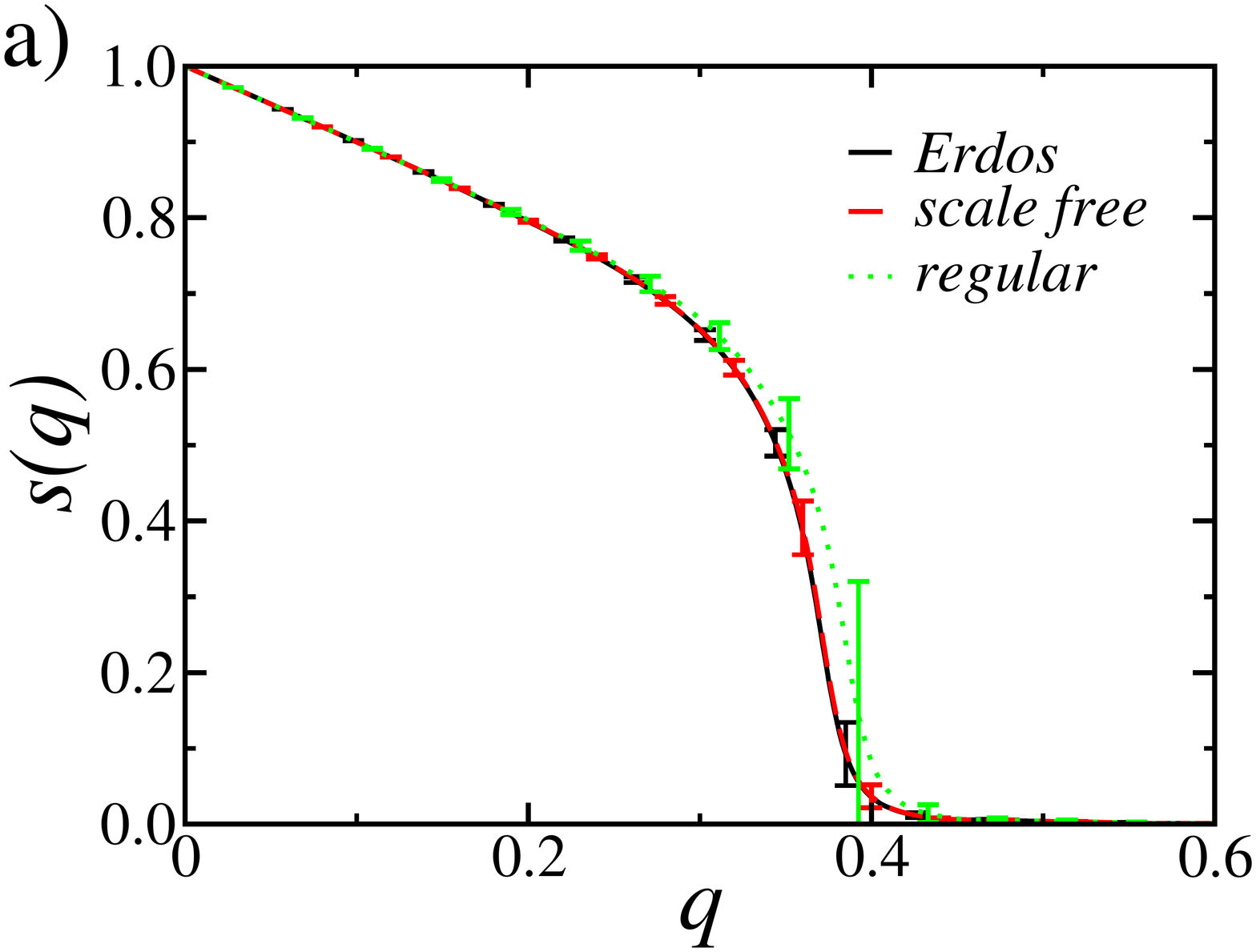}
\includegraphics[width=6.cm,angle = -90]{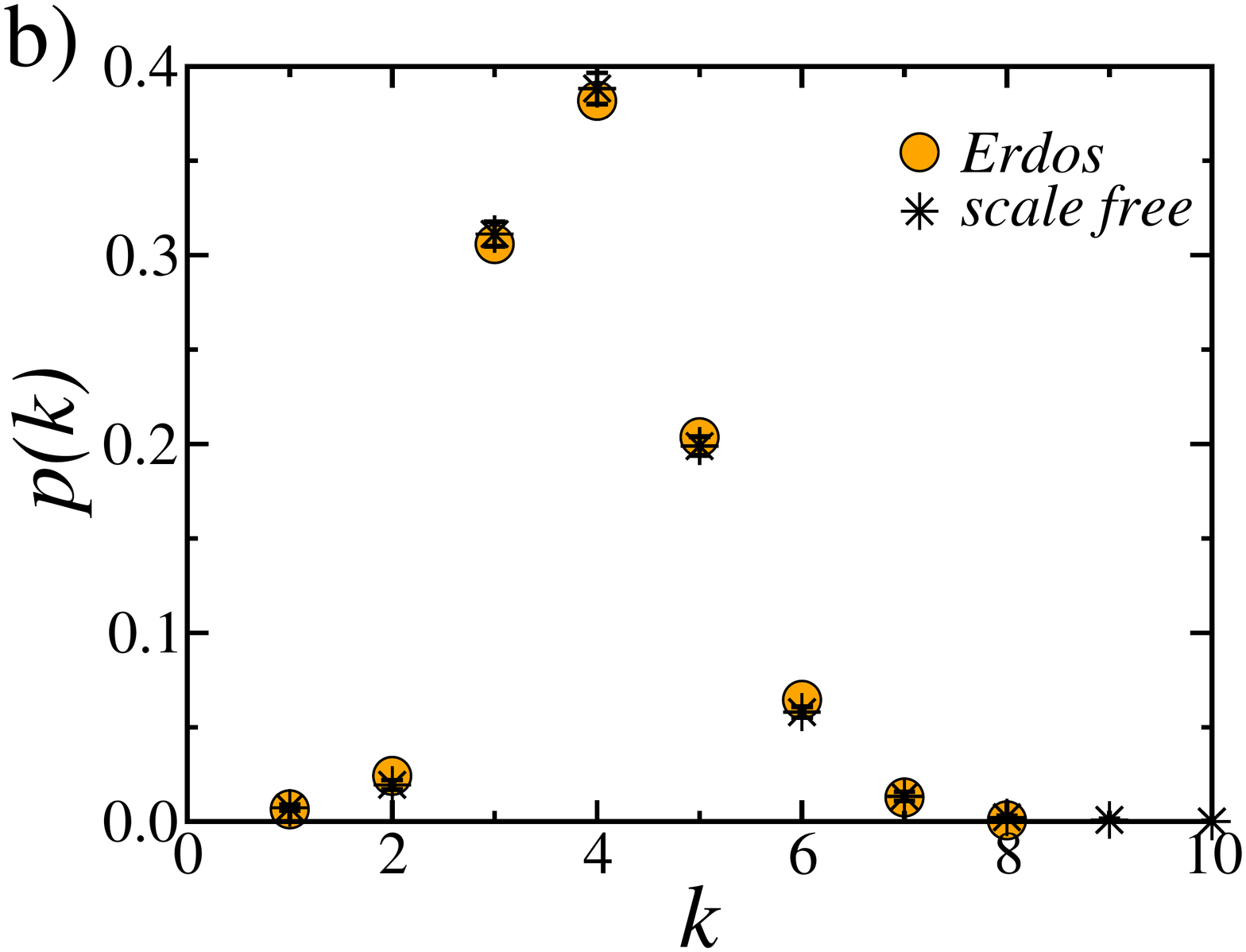}
\caption{a) The size of the largest connected cluster $s(q)$ versus the fraction of removed nodes for scale-free model networks, Erd\H{o}s-R\'enyi networks and random regular networks with $\langle k\rangle = 4$ after applying our algorithm. b) The degree distribution $p(k)$ versus the degree $k$ of these networks after applying our algorithm.}
\end{figure*}

\begin{figure*}[h]
\includegraphics[width=6.cm,angle = -90]{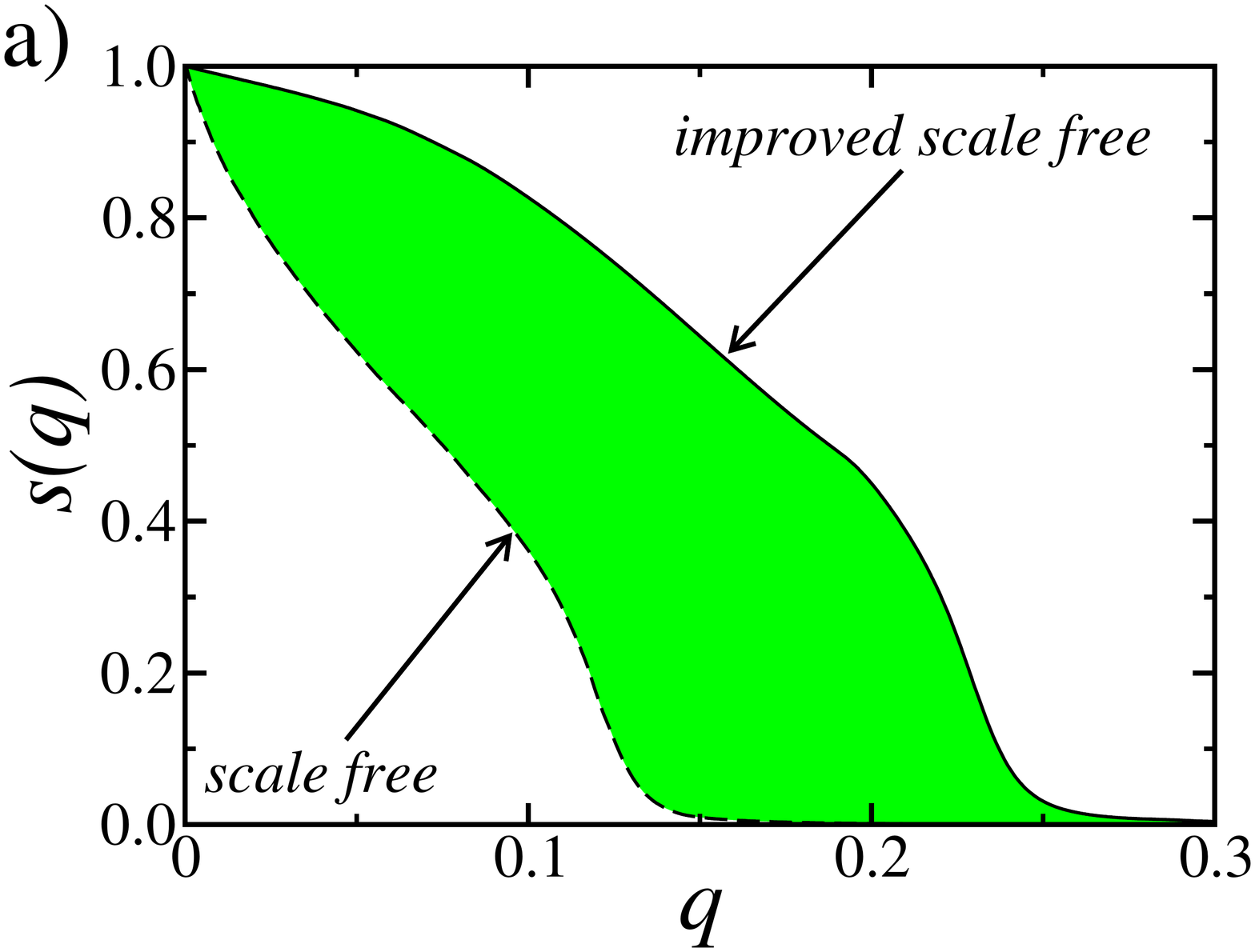}
\includegraphics[width=6.cm,angle = -90]{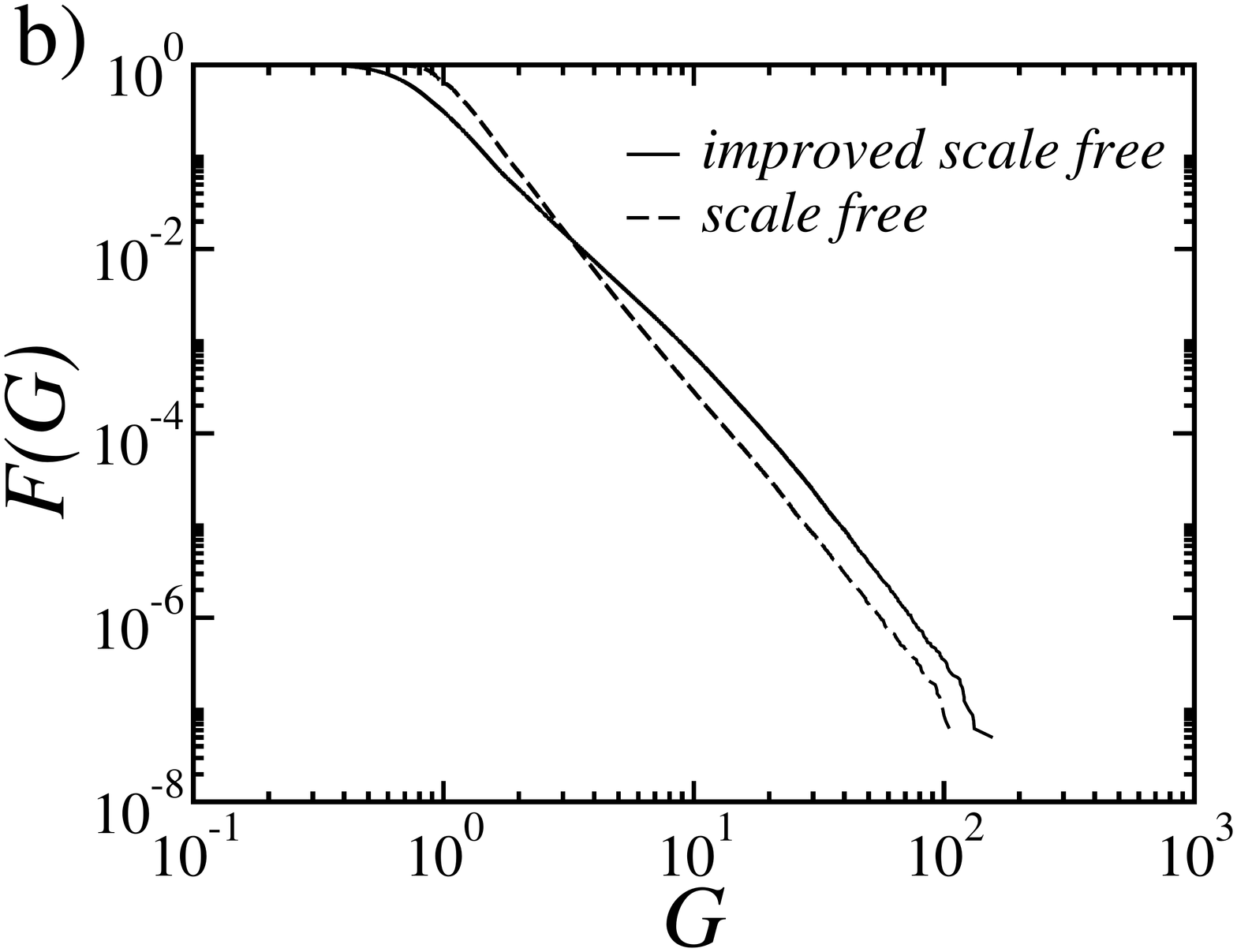}\\
\includegraphics[width=6.cm,angle = -90]{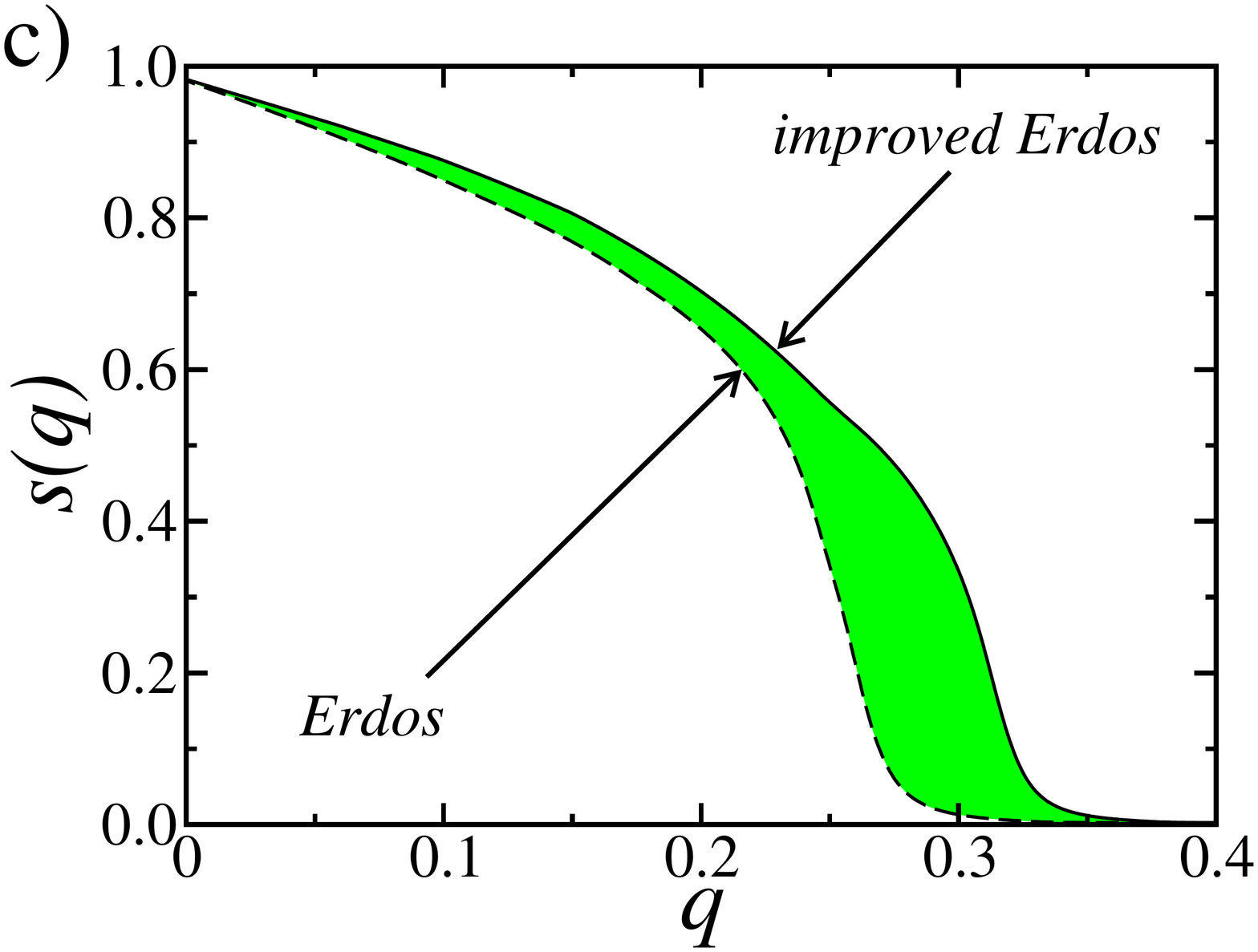}
\includegraphics[width=6.cm,angle = -90]{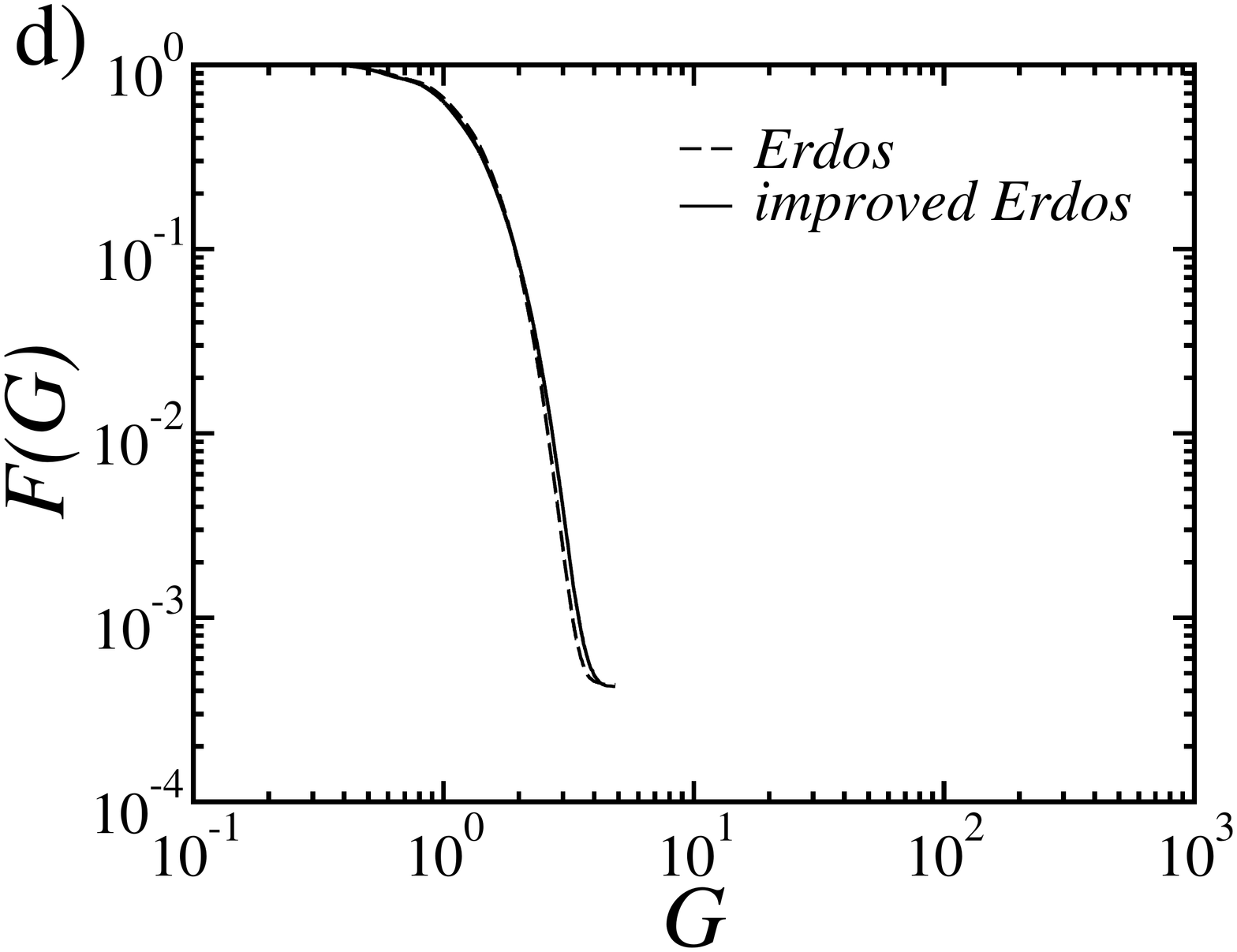}
\caption{In (a) we show the fraction of sites belonging to the largest connected cluster $s(q)$ versus the number of attacked nodes $q$ before and after applying our algorithm on scale-free networks with $N = 4000$ and $\gamma = 2.5$. The same is shown in (c) for Erd\H{o}s-R\'enyi networks with $N = 4000$ and $M = 8000$. The cumulative conductance distributions $F(G)$ versus the conductance $G$ for scale-free and Erd\H{o}s-R\'enyi networks are shown in (b) and (d), respectively, before and after the changes. Conductances between two nodes are measured between all pairs of nodes, assuming that each link in the network has unitary conductance.}
\end{figure*}

\begin{figure*}[h]
 \includegraphics[width=6cm,angle=-90]{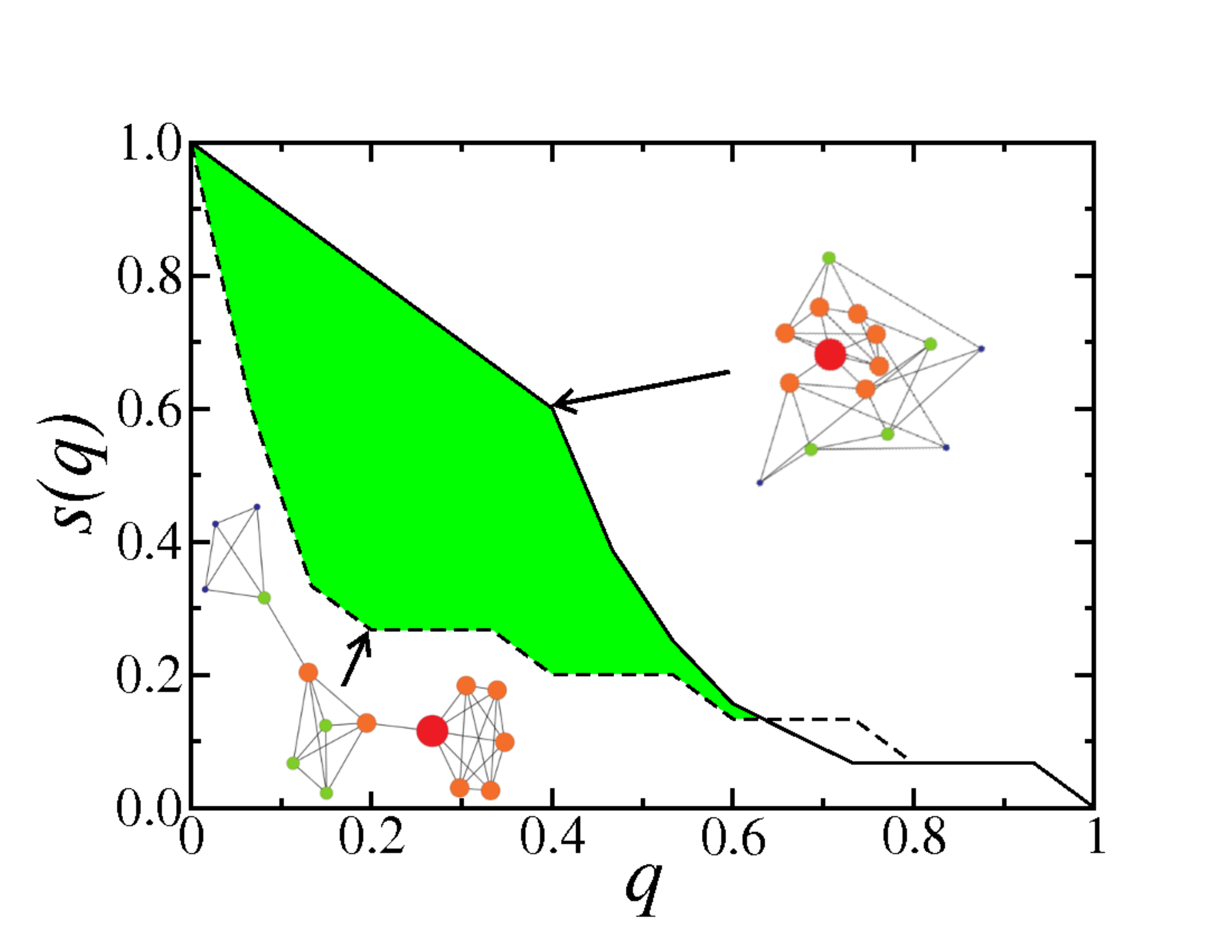}
\caption{The size of the largest connected cluster $s(q)$ versus the fraction of removed nodes for a highly assortative network(dashed line) and its improved onion-like counterpart(straight line). The figure also show the different structures of both networks.}
\end{figure*}

\begin{figure*}[h]
\includegraphics[width=6.cm,angle = -90]{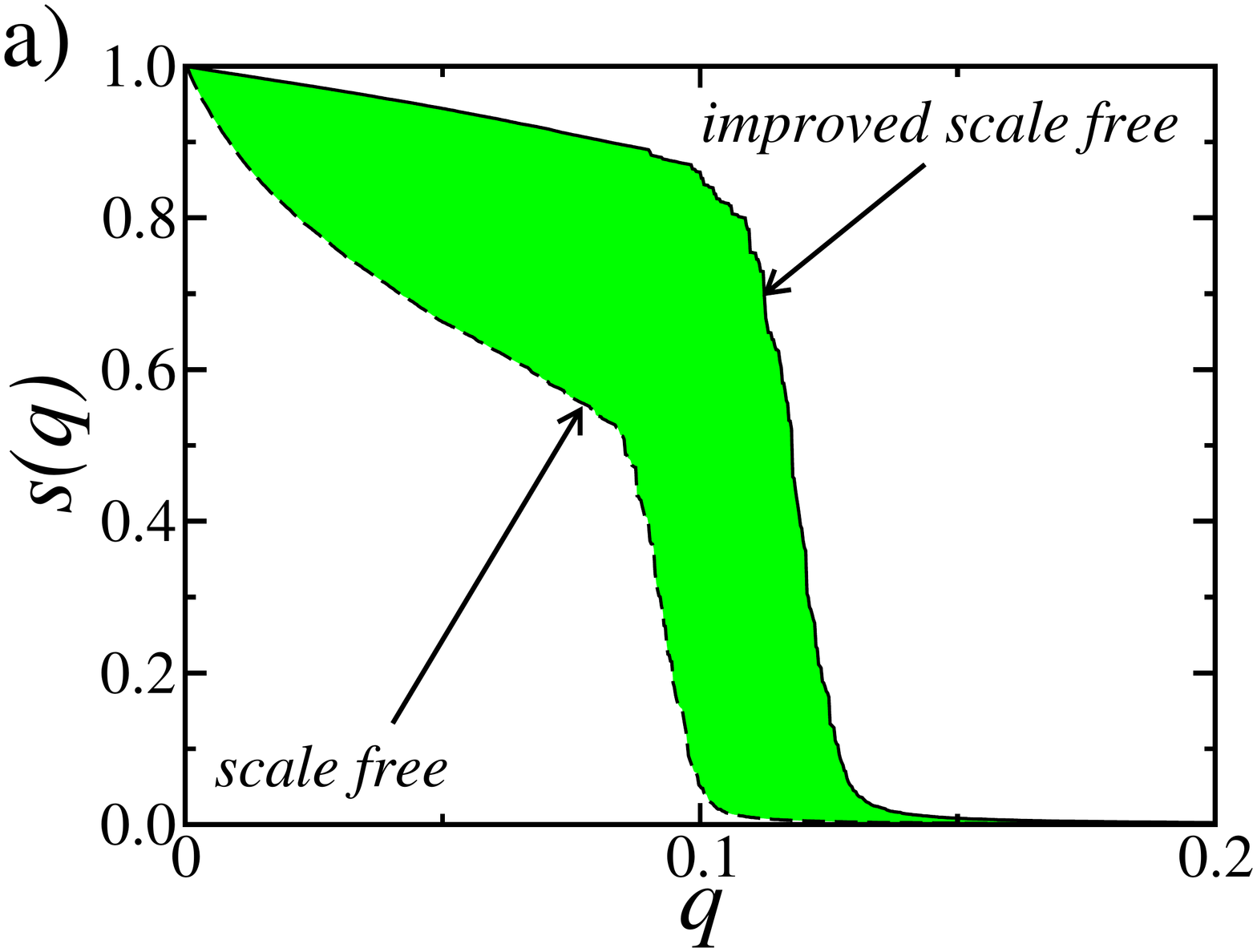}
\includegraphics[width=6.cm,angle = -90]{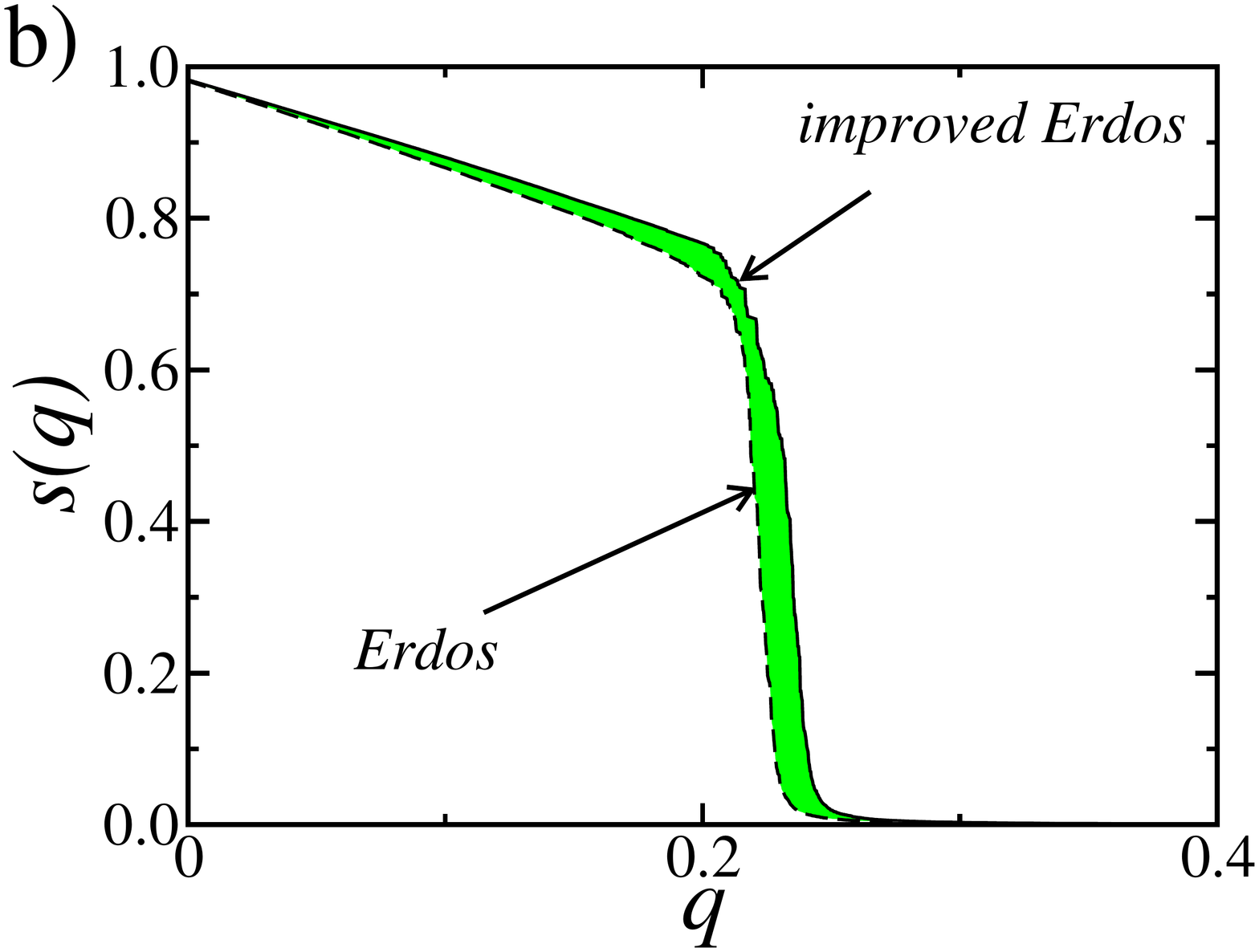}
\caption{The fraction of sites that belong to the largest connected cluster $s(q)$ vs the fraction of attacked nodes $q$ before and after applying our algorithm using the high betweenness adaptive attack for (a) scale-free networks with $N = 4000$ and $\gamma = 2.5$ and (b) for Erd\H{o}s-R\'enyi networks with $N = 4000$ and $M = 8000$. The high betweenness adaptive attack is the most effective malicious attack, but it requires global network information.}
\end{figure*}

\begin{figure*}[h]
\includegraphics[width=6.cm,angle = -90]{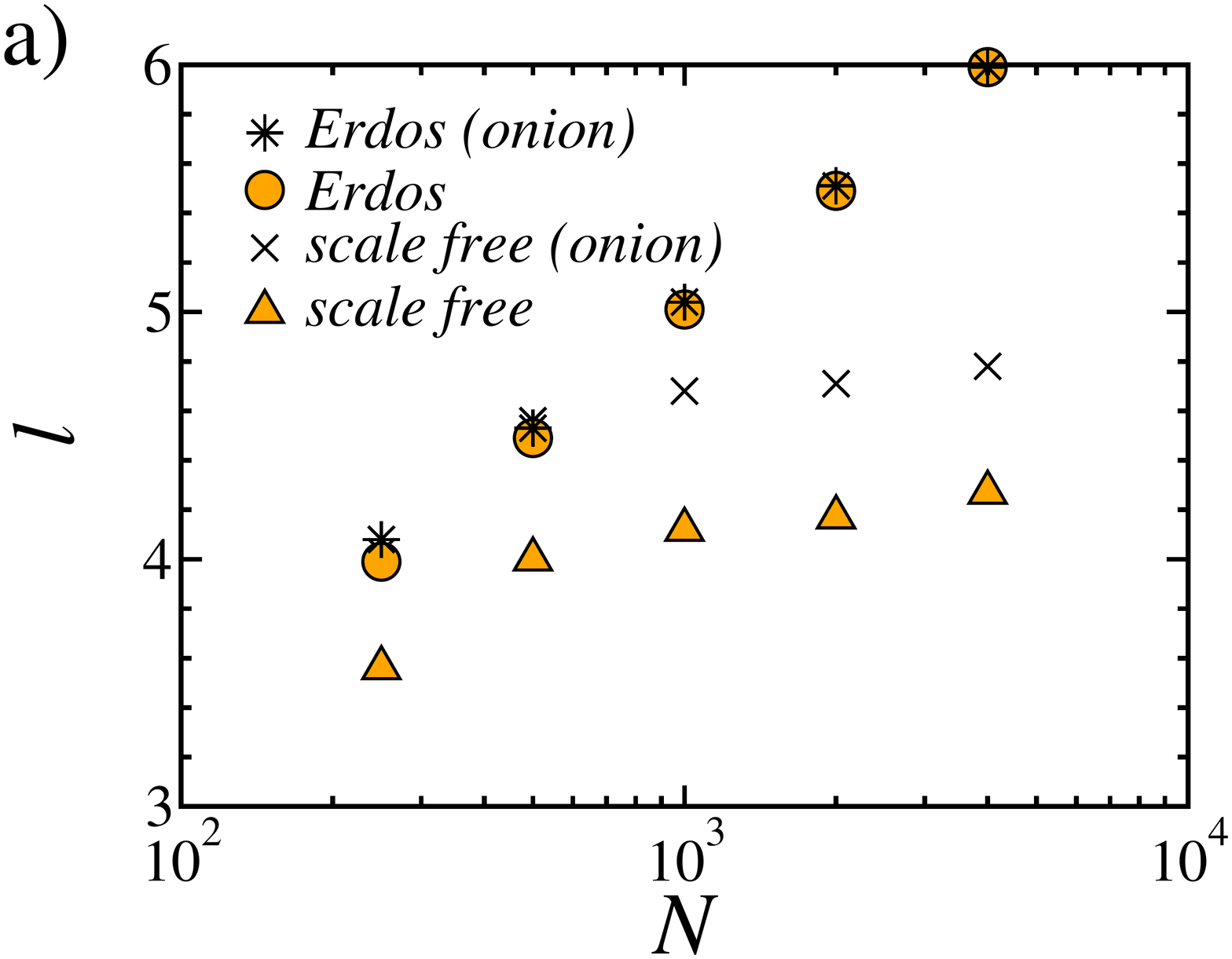}
\includegraphics[width=6.cm,angle = -90]{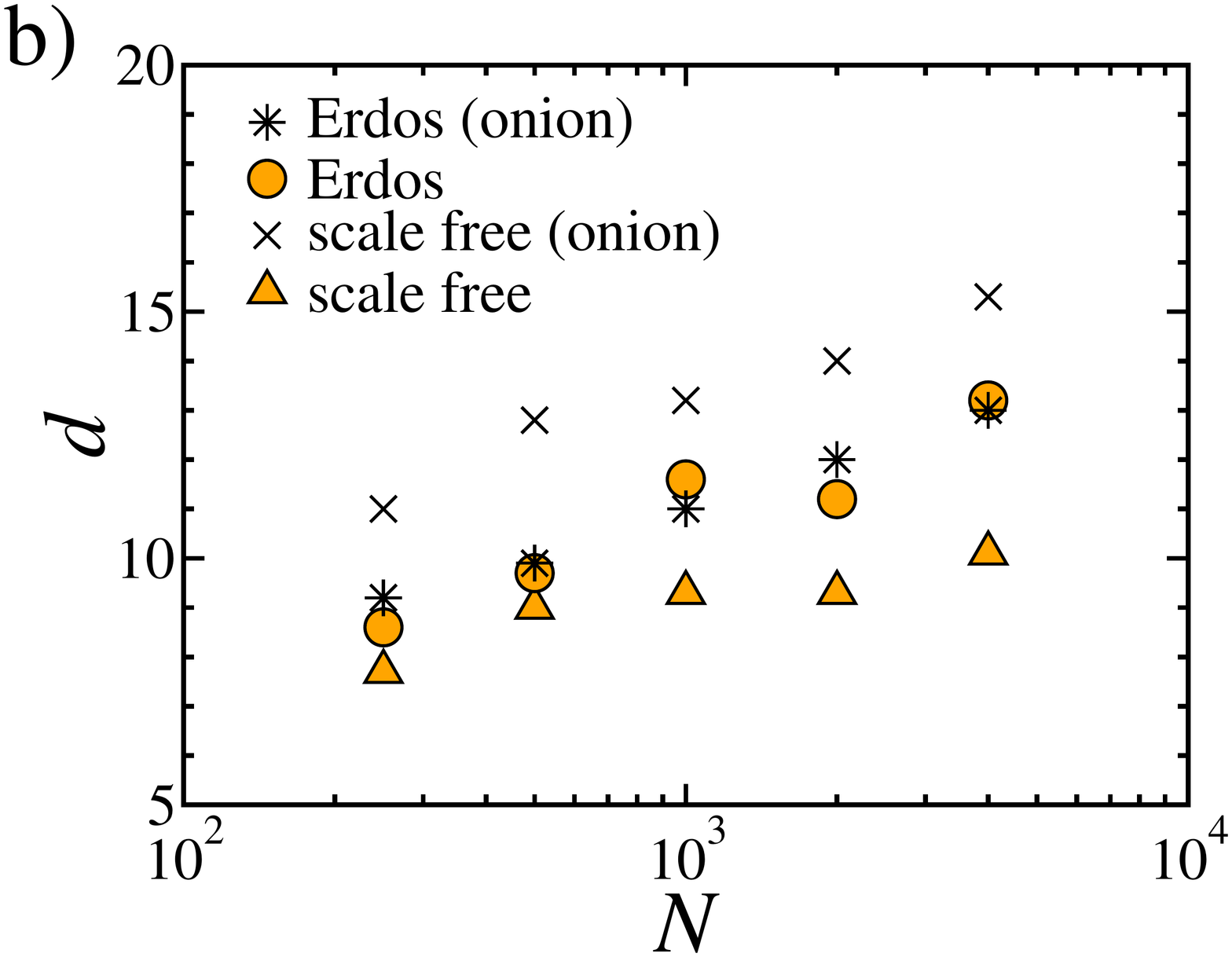}
\caption{The (a) average shortest path length $l$ and (b) diameter $d$ vs system size $N$ before and after applying our algorithm on scale-free networks with $\gamma = 2.5$ and on Erd\H{o}s-R\'enyi networks with $\langle k \rangle = 4$.}
\end{figure*}

\end{document}